\documentclass[aps,prl,twocolumn,showpacs,groupedaddress]{revtex4}

\usepackage{graphicx}
\usepackage{color}

\begin{document}

\preprint{}

\title{Atomically flat interface between a single-terminated LaAlO$_{3}$ substrate and SrTiO$_{3}$ thin film is insulating}

\author{Z. Q. Liu$^{1,2}$}

\author{Z. Huang$^{1}$}

\author{W. M. L\"{u}$^{1,3}$}

\author{K. Gopinadhan$^{1,3}$}

\author{X. Wang$^{1,2}$}

\author{A. Annadi$^{1,2}$}

\author{T. Venkatesan$^{1,2,3}$}

\author{Ariando$^{1,2}$}

\altaffiliation[Email: ]{ariando@nus.edu.sg}

\affiliation{$^1$NUSNNI-Nanocore, National University of Singapore,
5A Engineering Drive 1, Singapore 1174111 $^2$Department of Physics,
National University of Singapore, 2 Science Drive 3, Singapore
117542 $^3$Department of Electrical and Computer Engineering,
National University of Singapore, Singapore 117576}

\date{\today}

\begin{abstract}
The surface termination of (100)-oriented LaAlO$_{3}$ (LAO) single
crystals was examined and the growth of SrTiO$_{3}$ (STO) films on
LAO substrates was investigated. Eventually, the atomically flat
interface between a single-terminated LAO substrate and STO film was
achieved, which is expected to be similar to the \emph{n}-type
interface of two-dimensional electron gas
(2DEG),\emph{i.e.},(LaO)-(TiO$_{2}$). Particularly, that can serve
as a mirror structure for the typical 2DEG heterostructure to
further investigate the origin of 2DEG. This newly developed
interface was determined to be highly insulating and remained
insulating even after 1 h vacuum reduction. Additionally, this study
demonstrates an approach to achieve atomically flat film growth
based on LAO substrates.
\end{abstract}



\maketitle

The high mobility 2DEG generated at the atomically flat interface
between LAO films and TiO$_{2}$-terminated STO substrates was
reported for the first time by Ohtomo and Hwang.
\textcolor[rgb]{0.00,0.00,1.00}{$^{1}$} Since then, 2DEG at the
LAO/STO interface has attracted great attention from the community
of oxide electronics and been one of the most fascinating topics in
the contemporary condensed matter physics. Many intriguing physical
properties have been unveiled at such interface by different groups
, \emph{e.g.}, the electronically coupled complementary
interfaces,\textcolor[rgb]{0.00,0.00,1.00}{$^{2}$} the transition
from metallic to insulating phase as the thickness of LAO films
decreases to less than 4 unit cells
(uc),\textcolor[rgb]{0.00,0.00,1.00}{$^{3}$} the magnetic Kondo
scattering at the interface,\textcolor[rgb]{0.00,0.00,1.00}{$^{4}$}
the low temperature two-dimensional
superconductivity\textcolor[rgb]{0.00,0.00,1.00}{$^{5}$} and the
large gate capacitance enhancement due to strong electron
correlations.\textcolor[rgb]{0.00,0.00,1.00}{$^{6}$} Intriguingly,
Ariando \emph{et al.}\textcolor[rgb]{0.00,0.00,1.00}{$^{7}$}
recently found the coexistence of magnetic and superconducting-like
state and proposed electronic phase separation for the 2DEG at the
LAO/STO interface. Later the coexistence of superconductivity and
ferromagnetism in this system was observed by Dikin \emph{et
al.}\textcolor[rgb]{0.00,0.00,1.00}{$^{8}$} and Lu Li \emph{et
al.}.\textcolor[rgb]{0.00,0.00,1.00}{$^{9}$} The theoretical origin
of this coexistence is attributed to the Ti 3\emph{d} interface
electrons.\textcolor[rgb]{0.00,0.00,1.00}{$^{7,10}$} Moreover, the
integration of 2DEG with Si recently demonstrated by Park \emph{et
al.}\textcolor[rgb]{0.00,0.00,1.00}{$^{11}$} has pushed this
functional oxide interface system forward to nanoelectronic device
application.

As to the origin of the metallic conductivity at the LAO/STO
interface, generally three different mechanisms have been proposed.
The most possible mechanism is the polar discontinuity induced
interface charge transfer to overcome polarization catastrophe owing
to the polar and nonpolar nature of LAO and STO,
respectively.\textcolor[rgb]{0.00,0.00,1.00}{$^{12}$} The other
possible mechanisms are intermixing of La and Ti atoms at the
interface\textcolor[rgb]{0.00,0.00,1.00}{$^{13,14}$} and induced
oxygen vacancies\textcolor[rgb]{0.00,0.00,1.00}{$^{15,16}$} on the
STO side due to the oxygen growth pressure or the bombardment during
the growth process,\textcolor[rgb]{0.00,0.00,1.00}{$^{17}$} which
could behave similarly as in thermally reduced bulk STO (Ref.
\textcolor[rgb]{0.00,0.00,1.00}{18}) and thus lead to metallicity.
Additionally, recent studies
\textcolor[rgb]{0.00,0.00,1.00}{$^{19}$} indicate that the epitaxial
strain at the interface seems to play an important role in the
interface conductivity.

STO is a widely used substrate for atomically flat interface
engineering not only because of its excellent chemical and thermal
stabilities and the lattice match with other perovskite oxides, but
also the atomically controllable surface
termination.\textcolor[rgb]{0.00,0.00,1.00}{$^{20}$} Similar to STO,
LAO is also an excellent substrate and extensively utilized for
oxide thin film growth. Moreover, it is extremely difficult to
generate conductivity in LAO by thermal treatment because the
diffusion coefficient of oxygen vacancies in LAO is very
low.\textcolor[rgb]{0.00,0.00,1.00}{$^{21}$} In addition, oxygen
vacancies in LAO, if there is any, are energetically favored to be
in energy levels of $\sim2$ eV below the conduction
band,\textcolor[rgb]{0.00,0.00,1.00}{$^{22}$} which are too low and
nearly impossible to serve as a donor level at room temperature.
This is well in contrast to STO, in which the defect level of oxygen
vacancies could only be several ten meV below the conduction
band,\textcolor[rgb]{0.00,0.00,1.00}{$^{23}$} and hence gives an
advantage over STO for high temperature oxide film deposition.

The surface termination of (100)-oriented LAO single crystals has
been previously investigated by Yao \emph{et
al.}\textcolor[rgb]{0.00,0.00,1.00}{$^{24}$} and Wang \emph{et
al.}\textcolor[rgb]{0.00,0.00,1.00}{$^{25}$} with multiple
surface-sensitive techniques,\emph{e.g.}, time-of-flight scattering
and recoiling spectrometry, atomic force microscopy (AFM),
low-energy electron diffraction, Auger electron spectroscopy, x-ray
photoelectron spectroscopy and reflection electron microscopy. They
found that the surface termination of a LAO single crystal is
strongly temperature-dependent: the surface is exclusively
terminated in a Al-O layer from room temperature up to $\sim150$
$^\circ$C and a La-O layer above $\sim250$ $^\circ$C; Only in the
intermediate temperature region 150 $^\circ$C and 250 $^\circ$C,
mixed terminations exist. Using AFM
\textcolor[rgb]{0.00,0.00,1.00}{$^{24}$} they also demonstrated an
atomically flat LAO surface with uniform one-unit-cell step flow,
which actually provides the possibility to prepare an atomically
flat interface on a LAO substrate. More importantly, the realization
and investigation of an atomically flat interface between a LAO
substrate and STO film can provide a mirror structure for the
typical 2DEG heterostructure, which could be useful for
investigating the origin of 2DEG.

\begin{figure}
\includegraphics[width=3.4in]{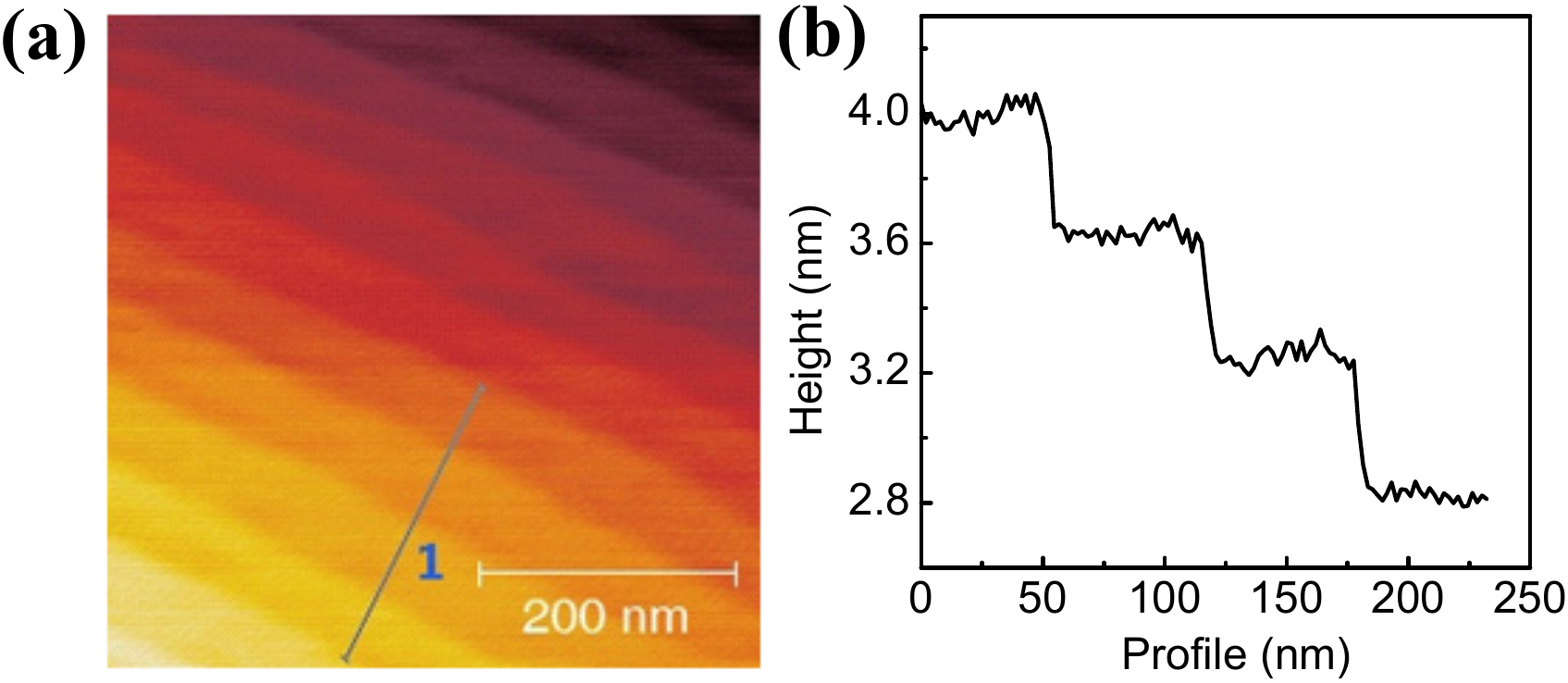}
\caption{\label{fig1} (a) Room temperature atomic force microscopic
image of a 0.5 $\mu$m $\times$ 0.5 $\mu$m area of a (100)-oriented
LaAlO$_{3}$ (LAO) single crystal annealed in air at 1000 $^\circ$C
for 2 hours. (b) The height profile of the cross line `1' in (a).}
\end{figure}

The as-received (100)-oriented LAO single crystals (CrysTec GmbH,
Germany) with two sides polished were examined by AFM at room
temperature, and indeed the steps on the surfaces were able to be
seen, which is consistent with the earlier
report.\textcolor[rgb]{0.00,0.00,1.00}{$^{24}$} Moreover, the
surface topography of LAO was largely improved by the subsequent air
annealing at 1000 $^\circ$C for 2 hours. The AFM image of an
annealed LAO single crystal is shown in Fig.
\textcolor[rgb]{0.00,0.00,1.00}{1(a)}. As can be seen, the surface
is single-terminated and atomically flat with uniform steps. The
profile data in Fig. \textcolor[rgb]{0.00,0.00,1.00}{1(b)}
corresponding to the cross line mark by `1' in Fig.
\textcolor[rgb]{0.00,0.00,1.00}{1(a)} displays the average terrace
width of $\sim60$ nm and the average step height of $\sim4$ {\AA}
pertaining to 1 uc of LAO.

\begin{figure}
\includegraphics[width=3.4in]{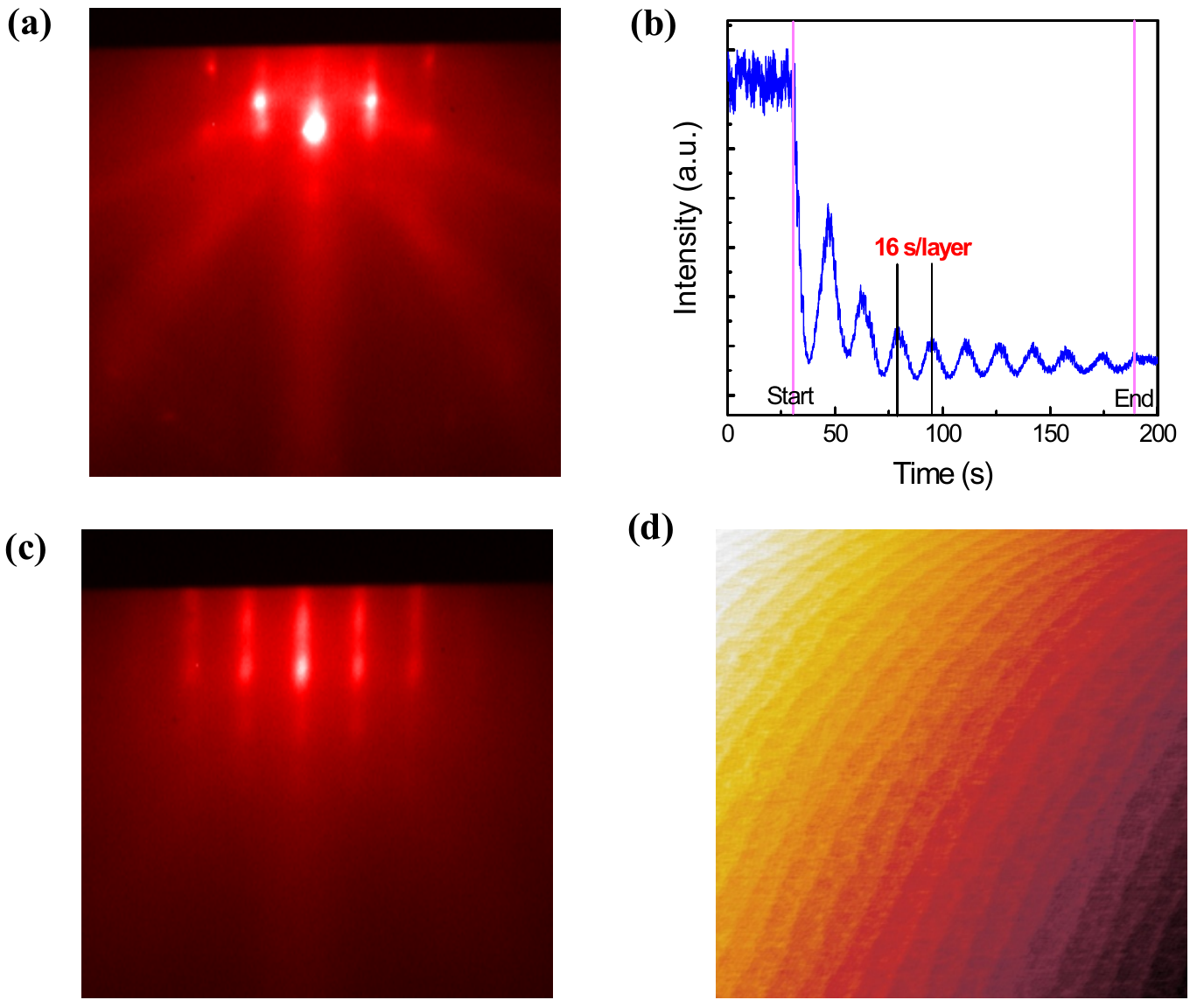}
\caption{\label{fig2} (a) Reflection high-energy electron
diffraction (RHEED) pattern of a LAO single crystal at 800 $^\circ$C
before deposition. (b) RHEED oscillations of a 10 unit cells (uc)
SrTiO$_{3}$ (STO) film grown on a fully-terminated LAO substrate.
(c) RHEED pattern after 10 uc STO film deposition. (d) AFM image of
an as-deposited 10 uc STO fil (1 $\mu$m$\times$1$\mu$m }
\end{figure}

In our work, pulsed laser deposition with a KrF excimer laser
($\lambda = 248$ nm) was used. STO films were deposited from a
single crystal STO target on fully-terminated LAO single crystal
substrates at 800 $^\circ$C and oxygen pressure of $5\times10^{-4}$
mbar, which are typical conditions for 2DEG growth. Before
deposition, LAO substrates were kept under the deposition conditions
for 30 mins for the stabilization of surface termination. The
reflection high-energy electron diffraction (RHEED) was utilized to
monitor the entire growth process. The RHEED pattern of a LAO
substrate at 800 $^\circ$C before deposition is shown in Fig.
\textcolor[rgb]{0.00,0.00,1.00}{2(a)}. During the deposition, the
fluence of laser energy was 1 J/cm$^{2}$ and the frequency was 1 Hz.
Consequently, the layer-by-layer growth of STO films on LAO
substrates was achieved. In Fig.
\textcolor[rgb]{0.00,0.00,1.00}{2(b)}, the RHEED oscillation of a 10
uc STO is shown with an accurate periodicity of 16 s/uc, indicating
a high quality layer-by-layer growth. The RHEED pattern after 10 uc
STO film deposition is shown in Fig.
\textcolor[rgb]{0.00,0.00,1.00}{2(c)}. The surface of the 10 uc STO
sample was examined by AFM and the terraces can be seen in Fig. 2(d)
with analogous signatures as the previous single-terminated LAO
substrate. Similarly, 25 uc STO films were also layer-by-layer grown
on a single-terminated LAO substrate.

Upon heating to the deposition temperature of 800 $^\circ$C, the
surface termination of a LAO substrate will experience a transition
from a Al-O layer at room temperature to a La-O layer possibly due
to the creation of surface oxygen deficiencies during
heating.\textcolor[rgb]{0.00,0.00,1.00}{$^{24}$} Therefore, the
(LaO)-(TiO$_{2}$) interface is expected to form in our case, which
is similar to the \emph{n}-type interface as in the typical 2DEG
LAO/STO heterostructure.\textcolor[rgb]{0.00,0.00,1.00}{$^{12}$}
Subsequently, the transport properties of both the 10 and 25 uc STO
samples were examined by simple four-probe linear DC resistance
measurements with a typical distance between voltage electrodes 1
mm. The samples were contacted using Al wires and the measuring
current was 5 nA. Nevertheless, it was found that both samples are
highly insulating with a resistance on the order of G$\Omega$ at
room temperature, in contrast to the metallicity of 2DEG observed in
the case of \emph{n}-type LAO/STO interface. The temperature
dependence of resistance (\emph{R-T}) for the 25 uc STO sample is
shown in Fig. \textcolor[rgb]{0.00,0.00,1.00}{3(a)}, which shows a
characteristic behavior of insulators.

\begin{figure}
\includegraphics[width=3.4in]{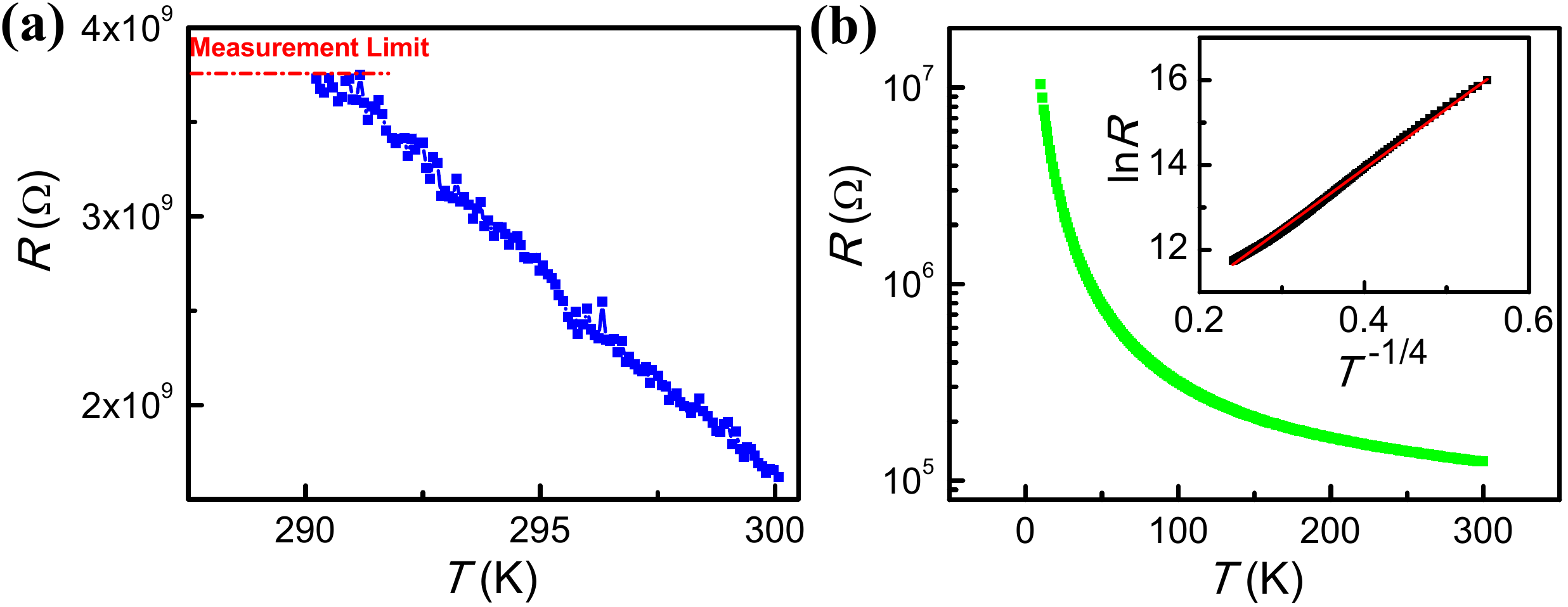}
\caption{\label{fig3}(a) Temperature dependence of the resistance
(\emph{R-T}) of a heterostructure with 25 uc STO film grown on a
single-terminated LAO substrate measured by 5 nA. (b) \emph{R-T}
curve of the heterostructure after 1 hour thermal reduction at 950
$^\circ$C and 4 $\times$ 10$^{-7}$ mbar vacuum measured by 1 $\mu$A
with completely the same geometry of (a). (Inset) Fitting plot in
terms of variable-range hopping of Mott law.}
\end{figure}

Microscopically the internal atomic layers of a LAO single crystal,
\emph{i.e.}, AlO$_{2}$- and LaO-layers, are polar, but the surface
of a bulk LAO single crystal would likely experience surface
reconstruction to be nonpolar. Thus, the growth of STO films on LAO
substrates can be explained by a scenario that a nonpolar material
is growing on another nonpolar surface. Therefore, the nonexistence
of polar discontinuity at the interface between STO films and LAO
substrates could be responsible for the highly insulating behavior.
On the other hand, the epitaxial interface strain for STO films
grown on LAO substrates works in the opposite direction compared to
the case of LAO films grown on STO substrates, which could also
affect the interface conductivity and hence account for the highly
insulating behavior. However, the nanoscale intermixing of the
interface atoms\textcolor[rgb]{0.00,0.00,1.00}{$^{13}$} seems not
suitable to clarify the highly insulating behavior here.

To investigate the effect of oxygen vacancies in the STO film on the
transport properties of this heterostructure, the 25 uc STO sample
was thermally reduced in a vacuum of 4 $\times$ 10$^{-7}$ mbar at
950 $^\circ$C for 1 h. The resistance measurement was then performed
with the same measurement geometry as for Fig.
\textcolor[rgb]{0.00,0.00,1.00}{3(a)}. The \emph{R-T} curve from 300
to 10 K as shown on a semi-logarithmic scale in Fig.
\textcolor[rgb]{0.00,0.00,1.00}{3(b)} still represents an insulating
behavior although the room temperature resistance decreases by
nearly four orders of magnitude relative to the initial resistance
(after the reduction, the LAO substrate was still extremely high
insulating, which was checked from the backside of the crystal). The
room temperature Hall measurement displays that the areal carrier
density is $\sim2.12\times10^{12}$ cm$^{-2}$, which is two orders of
magnitude smaller than the ideal areal carrier density of 2DEG
induced by the interface charge transfer, \emph{i.e.},
$3.28\times10^{14}$ cm$^{-2}$. Furthermore, it was found that the
\emph{R-T} curve can be well fitted by the variable-range hopping of
Mott law, \emph{i.e.}, $\ln$\emph{R} $\propto$ $T^{-1/4}$ , over the
whole temperature range as plotted in the inset of Fig.
\textcolor[rgb]{0.00,0.00,1.00}{3(b)}, which reveals that the
\emph{R-T} curve belongs to the three dimensional transport
property. Hence considering the thickness of 25 uc STO, the areal
carrier density is converted into a volume density to be
$2.17\times10^{18}$ cm$^{-3}$. The value is smaller than the
experimental Mott critical carrier density $\sim5\times10^{18}$
cm$^{-3}$ (Ref. \textcolor[rgb]{0.00,0.00,1.00}{23}) of the
metal-insulator transition for a reduced STO film, so this can also
explain why the film is still insulating even after high temperature
vacuum annealing.

However, the insulating behavior in reduced STO thin films is
completely different from those of reduced thicker films and bulk
crystals, which show metal-insulator transition and metallicity,
respectively \textcolor[rgb]{0.00,0.00,1.00}{$^{24}$}. This could be
due to the compensating defects like Ti vacancies and Sr vacancies
dominant in ultrathin STO films, which are likely created in the
high temperature and high vacuum reduction process.

Not only can STO films be layer-by-layer grown on a fully-terminated
LAO substrate, the high quality atomically flat growth can also be
obtained for some other perovskite materials like NdAlO$_{3}$ and
PrAlO$_{3}$. This demonstrates an approach to achieve atomically
flat interfaces based on LAO substrates, which are robust and nearly
free of oxygen-vacancy-induced conductivity.

In summary, we demonstrated atomically flat interfaces between STO
films and single-terminated LAO substrates. The transport
measurements displayed that this type of interface is highly
insulating and remains insulating even after 1 hour vacuum
reduction. The reason for that could be the surface reconstruction
of LAO single crystals or due to the interface epitaxial strain.
This work can provide a fundamental perspective for the oxide
electronics community. Besides, our work opens a way to achieve
atomically flat film growth based on LAO substrates. Furthermore,
the quasi-2DEG could even also be tailored probably by means of
vacuum reduction or Argon-ion milling after the realization of
atomically flat nanoscale film growth on LAO substrates.

\begin{acknowledgments}
We thank M. Huijben and J. Mannhart for discussions and the National
Research Foundation (NRF) Singapore under the Competitive Research
Programme `Tailoring Oxide Electronics by Atomic Control', NUS
cross-faculty grant and FRC for financial support.
\end{acknowledgments}


\end{document}